\documentclass[preprint,showpacs,preprintnumbers,amsmath,amssymb,aps]{revtex4}

\usepackage{graphicx}
\usepackage{dcolumn}
\usepackage{bm}

\def\lesssim{\ \raise.3ex\hbox{$<$}\kern-0.8em\lower.7ex\hbox{$\sim$}\ }
\def\gesim{\ \raise.3ex\hbox{$>$}\kern-0.8em\lower.7ex\hbox{$\sim$}\ }

\def\rnum#1{\expandafter{\romannumeral #1}} 
\def\Rnum#1{\uppercase\expandafter{\romannumeral #1}} 
\usepackage{ascmac}
\usepackage{amssymb}
\usepackage{amsmath}
\usepackage{txfonts}
\usepackage{graphicx}

\newcommand{\lef}{\left}
\newcommand{\rig}{\right}
\newcommand{\up}{\uparrow}
\newcommand{\down}{\downarrow}

\newcommand{\Om}{\Omega}

\newcommand{\vep}{\varepsilon}

\begin{document}
\title{Specific heat and effects of pairing fluctuations in the BCS-BEC crossover regime of an ultracold Fermi gas}
\author{Pieter van Wyk, Hiroyuki Tajima, Ryo Hanai, and Yoji Ohashi}
\affiliation{Department of Physics, Keio University, 3-14-1 Hiyoshi, Kohoku-ku, Yokohama 223-8522, Japan}
\date{\today}
\begin{abstract}
We investigate the specific heat at constant volume $C_V$ in the BCS (Bardeen-Cooper-Schrieffer)-BEC (Bose-Einstein condensation) crossover regime of an ultracold Fermi gas above the superfluid phase transition temperature $T_{\rm c}$. Within the framework of the strong-coupling theory developed by Nozi\`eres and Schmitt-Rink, we show that this thermodynamic quantity is sensitive to the stability of preformed Cooper pairs. That is, while $C_V(T\gesim T_{\rm c})$ in the unitary regime is remarkably enhanced by {\it metastable} preformed Cooper pairs or pairing fluctuations, it is well described by that of an ideal Bose gas of long-lived {\it stable} molecules in the strong-coupling BEC regime. Using these results, we identify the region where the system may be viewed as an almost ideal Bose gas of stable pairs, as well as the pseudogap regime where the system is dominated by metastable preformed Cooper pairs, in the phase diagram of an ultracold Fermi gas with respect to the strength of a pairing interaction and the temperature. We also show that the calculated specific heat agrees with the recent experiment on a $^6$Li unitary Fermi gas. Since the formation of preformed Cooper pairs is a crucial key in the BCS-BEC crossover phenomenon, our results would be helpful in considering how fluctuating preformed Cooper pairs appear in a Fermi gas, to eventually become stable, as one passes through the BCS-BEC crossover region.
\end{abstract}
\pacs{03.75.Ss, 03.75.-b, 03.70.+k}
\maketitle
%%%%%%%%%%%%%%%%%%%%%%%%%%%%%%%%%%%%%%%%%%%%%%%%%%%%%%%%%%%%%%%%%%%%%%%%%%%%%% 
\par
\section{Introduction}
\par
The formation of preformed Cooper pairs is a crucial key in considering the BCS (Bardeen-Cooper-Schrieffer)-BEC (Bose-Einstein condensation) crossover phenomenon\cite{Eagles,Leggett,NSR,Randeria,Melo,OhashiGriffin,Perali1,Giorgini,Bloch,Chen,Ketterle}, where the character of a Fermi superfluid continuously changes from the weak-coupling BCS type to the BEC of tightly bound molecules with increasing the strength of a pairing interaction. Since the realization of this crossover phenomenon in $^{40}$K\cite{Jin} and $^6$Li\cite{Zwierlein,Kinast,Jochim} Fermi gases, by using a Feshbach resonance\cite{Timmermans,Holland,Chin}, it has extensively been discussed both theoretically\cite{OhashiGriffin2,HuiHu3,Haussmann,Fukushima,HuiHu2,Tsuchiya,Levin2,Tsuchiya2,Kashimura,Bulgac,Tajima} and experimentally\cite{Chin2,Ketterle2,Kinast2,Chin3,Jin2,Thomas,Gaebler,Horikoshi,Salomon,Perali,Sanner,Ku,Jin3} how metastable preformed Cooper pairs (that are also referred to in the literature as pairing fluctuations) appear in a Fermi gas, to eventually become long-lived stable pairs, as one passes through the BCS-BEC crossover region above the superfluid phase transition temperature $T_{\rm c}$. Since the BCS-BEC crossover is also considered as a crucial key in the fields of high-$T_{\rm c}$ cuprates \cite{Chen,Perali2}, as well as iron based superconductors \cite{Shigeru}, elucidating strong-coupling properties of an ultracold Fermi gas in this regime would also contribute to the understanding of these strongly corrleated electron systems. 
\par
Although there is no clear phase boundary between the weak-coupling BCS regime and the strong-coupling BEC regime, it is still an interesting problem to physically identify the region where preformed Cooper pairs dominate over system properties. In this regard, we note that, when preformed Cooper pairs appear in a normal Fermi gas, single-particle Fermi excitations are expected to have a gap-like structure, reflecting their finite dissociation energy. This so-called preformed pair scenario has been discussed in high-$T_{\rm c}$ cuprates\cite{Yanase}, as a possible mechanism of the pseudogapped density of states observed in the under-doped regime of this strongly correlated electron system\cite{Shen,Lee,Renner,Tanaka}. In this field, the temperature $T^*$ below which the pseudogap appears in the density of states $\rho(\omega)$ is called the pseudogap temperature. Although $T^*$ is not accompanied by any phase transition, it is a useful characteristic temperature to distinguish between the normal Fermi liquid regime and the pseudogap regime in the phase diagram of high-$T_{\rm c}$ cuprates.
\par
In high-$T_{\rm c}$ cuprates, the validity of the preformed pair scenario is still in debate\cite{Yanase,Pines,Kamp,Chakravarty}, due to the complexity of this system. In contrast, since an ultracold Fermi gas in the BCS-BEC crossover region is simply dominated by strong pairing fluctuations, the preformed pair scenario is validated. Indeed, it has been pointed out\cite{Gaebler,Tsuchiya2} that the back-bending curve of the single-particle dispersion observed by a recent photoemission-type experiment on a $^{40}$K unitary Fermi gas\cite{Jin2} may be a signature of the pseudogap phenomenon. It has also been shown\cite{Kashimura,Tajima} that the anomalous suppression of the uniform spin susceptibility $\chi_{\rm s}$ observed in a $^6$Li Fermi gas above $T_{\rm c}$\cite{Sanner} can be explained as an effect of fluctuating spin-singlet preformed pairs. At present, although the pseudogap temperature $T^*$ has not experimentally been determined in an ultracold Fermi gas, the existence of this characteristic temperature has theoretically been predicted from the calculated density of states $\rho(\omega)$\cite{Tsuchiya}. As in the case of high-$T_{\rm c}$ cuprates, $T^*$ in an ultracold Fermi gas is not accompanied by any phase transition. However, it physically gives the boundary between a normal Fermi gas regime and the pseudogap regime, being dominated by fluctuating preformed Cooper pairs, in the phase diagram of an ultracold Fermi gas above $T_{\rm c}$.
\par
A similar characteristic temperature $T_{\rm s}$, called the spin-gap temperature, has also been predicted\cite{Tajima}. $T_{\rm s}$ is determined as the temperature below which the spin susceptibility $\chi_{\rm s}$ in the normal state is anomalously suppressed by spin-singlet preformed Cooper pairs. Although $T_{\rm s}$ is not exactly the same as the pseudogap temperature $T^*$ they have essentially the same background physics, and thus $T_{\rm s}$ also has the meaning of the boundary between the normal Fermi gas regime and the preformed-pair regime. We briefly note that this so-called spin-gap phenomenon\cite{Millis} has also been discussed in high-$T_{\rm c}$ cuprates\cite{Yasuoka,Kitaoka}. 
\par
Although $T^*$ and $T_{\rm s}$ conveniently give the boundary around which metastable preformed Cooper pairs start to dominate over the system in the BCS-BEC crossover region, they do not have any information about where these fluctuating preformed pairs become stable in the strong-coupling BEC regime. In determining this second boundary, however, the low-energy density of states $\rho(\omega\sim 0)$ (which gives the pseudogap temperature $T^*$), as well as the spin susceptibility $\chi_{\rm s}$ (which gives the spin gap temperature $T_{\rm s}$), are not useful, because both quantities almost vanish deep inside BEC regime where most Fermi atoms form spin-singlet bound molecules with a large binding energy.
\par
In this regard, the specific heat at constant volume $C_V$ is promising, because it is finite in the whole BCS-BEC crossover region. In addition, $C_V$ is sensitive to the quantum statistics of particles in the system, in the sense that, while $C_V$ exhibits a linear-temperature dependence in a Fermi gas, it increases with decreasing the temperature in an ideal Bose gas. Furthermore, the specific heat has recently become accessible in cold Fermi gas physics\cite{Ku}. Thus, the above-mentioned second boundary may be determined by using this thermodynamic quantity.
\par
The purpose of this paper is to theoretically confirm this expectation, to distinguish between the pseudogap regime, which is dominated by metastable preformed Cooper pairs, or pairing fluctuations, and the region that can be viewed as a gas of long-lived stable pairs, in the phase diagram of an ultracold Fermi gas. Including pairing fluctuations above $T_{\rm c}$ within the framework of the strong-coupling theory developed by Nozi\`eres and Schmitt-Rink\cite{NSR}, we show that the temperature dependence of the specific heat is very different in between the BCS-BEC crossover region and the strong-coupling BEC regime. Using this difference, we determine a characteristic temperature ${\tilde T}$ which conveniently gives the boundary between the pseudogap regime and the region of stable pairs. In addition, the specific heat is also shown to be able to determine the boundary between the normal Fermi gas regime and the pseudogap regime. The characteristic temperature ${\bar T}$ giving the latter boundary is found to be consistent with the previous pseudogap temperature $T^*$, as well as the spin-gap temperature $T_{\rm s}$, that are, respectively, obtained from the density of states $\rho(\omega)$ and the spin susceptibility $\chi_{\rm s}$. We also show that that our result on $C_V$ agrees with the recent experiment on a $^6$Li unitary Fermi gas\cite{Ku}. We briefly note that the specific heat in a unitary Fermi gas has also been discussed within a $T$-matrix approximation\cite{Perali}, as well as within the combined NSR theory with local density approximation\cite{HuiHu3}. 
\par
This paper is organized as follows. In Sec. II, we explain our strong coupling formalism used to calculate the specific heat at constant volume $C_V$ in the BCS-BEC crossover region above $T_{\rm c}$. In Sec. III, we show our numerical results on $C_V$ over the entire BCS-BEC crossover region. Here, we explain how to determine ${\tilde T}$ and ${\bar T}$ from the temperature dependence of $C_V$. Using these characteristic temperatures, we identify the region where the system is dominated by fluctuating metastable preformed Cooper pairs, as well as the region where the system is dominated by long-lived stable molecules, in the phase diagram of an ultracold Fermi gas with respect to the interaction strength and the temperature. Throughout this paper, we set $\hbar=k_{\rm B}=1$, and the system volume $V$ is taken to be unity, for simplicity.
\par
%%%%%%%%%%%%%%%%%%%%%%%%%%%%%%%%%%%%%%%%%%%%%%%%%%%%%%%%%%%%%%%%%%%%%%%%%%%%%%
\section{Formulation}
\par
We consider a two-component uniform Fermi gas in the normal state, described by the BCS Hamiltonian,
\begin{equation}
H=\sum_{{\bm p},\sigma}\xi_{\bm p}
c^\dagger_{{\bm p},\sigma}c_{{\bm p},\sigma}
-U\sum_{{\bm p},{\bm p}',{\bm q}}
c^\dagger_{{\bm p}+{\bm q}/2,\up}
c^\dagger_{-{\bm p}+{\bm q}/2,\down}
c_{-{\bm p}'+{\bm q}/2,\down}
c_{{\bm p}'+{\bm q}/2,\up},
\label{eq.1} 
\end{equation}   
where $c_{{\bm p},\sigma}$ is the annihilation operator of a Fermi atom with pseudospin $\sigma=\uparrow,\downarrow$, describing two atomic hyperfine states. $\xi_{\bm p}=\varepsilon_{\bm p}-\mu={\bm p}^2/(2m)-\mu$ is the kinetic energy, measured from the Fermi chemical potential $\mu$ (where $m$ is an atomic mass). $-U$ is an $s$-wave pairing interaction, which we treat as a tunable parameter. As usual, we measure the interaction strength in terms of the $s$-wave scattering length $a_s$, which is related to the pairing interaction $-U$ as,
\begin{equation}
{4\pi a_s \over m}
=
-{
U \over 1-U\sum_{\bm p}^{p_{\rm c}}{1 \over 2\vep_{\bm p}}
},
\label{eq.2}
\end{equation}
where $p_{\rm c}$ is a momentum cutoff.
\par
%%%%%%%%%%%%%%%%%%%%%%%%%%%%%%%%%%%%%%%%%%%%%%%%%%%%%%%%%%%%%%%%%%%%%%%%%%%%%%%
\begin{figure}[t]
\begin{center}
\includegraphics[width=15.0cm]{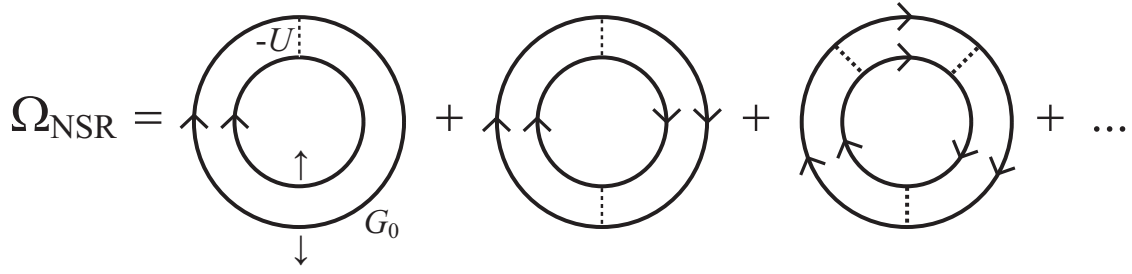}
\end{center}
\caption{Feynman diagrams describing the correction term $\Om_{\text{NSR}}$ to the thermodynamic potential $\Omega$ in the strong-coupling NSR theory\cite{NSR}. The solid line is the bare single-particle thermal Green's function $G_0^{-1}({\bm p},i\omega_n)=i\omega_n-\xi_{\bm p}$ (where $\omega_n$ is the fermion Matsubara frequency), and the dashed line is the attractive pairing interaction $-U$.}
\label{fig1}
\end{figure} 
%%%%%%%%%%%%%%%%%%%%%%%%%%%%%%%%%%%%%%%%%%%%%%%%%%%%%%%%%%%%%%%%%%%%%%%%%%%%%%%
\par
We include pairing fluctuations within the ordinary NSR theory\cite{NSR}. In this BCS-BEC crossover theory, the thermodynamic potential $\Omega=\Omega_0+\Omega_{\rm NSR}$ consists of the non-interacting part,
\begin{equation}
\Omega_0=-2T\sum_{\bm p}\ln\lef[1+e^{-\xi_{\bm p}/T}\rig],
\label{eq.3}
\end{equation}
and the fluctuation correction $\Omega_{\rm NSR}$, the latter of which is diagrammatically given in Fig. \ref{fig1}. The summation of these diagrams gives
\begin{equation}
\Omega_{\rm NSR} = -T\sum_{{\bm q},i\nu_n}
\ln\Gamma({\bm q},i\nu_n),
\label{eq.4} 
\end{equation}
where $\nu_n$ is the boson Matsubara frequency, and
\begin{equation}
\Gamma({\bm q},i\nu_n)=
{1 \over 
\displaystyle
{m \over 4\pi a_s}+
\left[
\Pi({\bm q},i\nu_n)-\sum_{\bm p}{1 \over 2\varepsilon_{\bm p}}
\right]
}				      
\label{eq.5}
\end{equation}
is the NSR particle-particle scattering matrix. Here,
\begin{equation}
\Pi({\bm q},i\nu_n)=
-\sum_{\bm p}
{1-f(\xi_{{\bm p}+{\bm q}/2})-f(\xi_{-{\bm p}+{\bm q}/2})
\over
i\nu_n-\xi_{{\bm p}+{\bm q}/2}-\xi_{-{\bm p}+{\bm q}/2}
}
\label{eq.6}
\end{equation}   
is the lowest-order pair-correlation function, describing fluctuations in the Cooper channel (where $f(x)$ is the Fermi distribution function).
\par
We calculate the specific heat at constant volume $C_V$ from the formula,
\begin{equation}
C_V=\lef(\frac{\partial E}{\partial T}\rig)_{V,N}.
\label{eq.7} 
\end{equation}
Here, the internal energy $E$ is obtained from $\Omega=\Omega_0+\Omega_{\rm NSR}$ via the Legendre transformation,
\begin{eqnarray}
E&=&\Omega-T\lef(\frac{\partial\Om}{\partial{T}}\rig)_{\mu}-\mu\lef(\frac{\partial\Om}{\partial{\mu}}\rig)_{T}
\nonumber\\
&=&
2\sum_{\bm p}\vep_{\bm p}f(\xi_{\bm p})
-T\sum_{{\bm q},i\nu_n}
\Gamma({\bm q},i\nu_n)
\left[
T\frac{\partial}{\partial{T}}\Pi({\bm p},i\nu_n)
+
\mu\frac{\partial}{\partial{\mu}}\Pi({\bm q},i\nu_n)
\right].
\label{eq.8} 
\end{eqnarray}
The Fermi chemical potential $\mu$ in Eq. (\ref{eq.8}) is determined from the equation for the total number $N$ of Fermi atoms, given by
\begin{eqnarray}
N
&=&-\left(\frac{\partial\Omega}{\partial\mu}\right)_{T}
\nonumber
\\
&=&2\sum_{\bm p}f(\xi_{\bm p})
-T\sum_{{\bm q},i\nu_n}
\Gamma({\bm q},i\nu_n)
\frac{\partial}{\partial\mu}\Pi({\bm q},i\nu_n)
\nonumber
\\
&=&
N_{\rm F}^0+
N_{\rm NSR},
\label{eq.9} 
\end{eqnarray}
where $N_{\rm F}^0$ and $N_{\rm NSR}$ represent the non-interacting part and the NSR strong-coupling corrections, respectively. In this paper, we numerically evaluate Eq. (\ref{eq.7}) from the internal energies $E(T)$ and $E(T+\delta T)$. 
\par
In the NSR theory\cite{NSR,Melo,Randeria,OhashiGriffin}, the equation for the superfluid phase transition temperature $T_{\rm c}$ is conveniently obtained from the Thouless criterion\cite{Thouless}, stating that the superfluid instability occurs when the particle-particle scattering matrix $\Gamma({\bm q},i\nu_n)$ in Eq. (\ref{eq.5}) has a pole at ${\bm q}=\nu_n=0$. The resulting $T_{\rm c}$-equation has the same form as the mean-field BCS gap equation at $T_{\rm c}$, as
\begin{equation}
1=-{4\pi a_s \over m}
\sum_{\bm p}
\left[
{1 \over \xi_{\bm p}}\tanh{\xi_{\bm p} \over 2T}
-{1 \over 2\varepsilon_{\bm p}}
\right].
\label{eq.10}
\end{equation}
Following the standard NSR approach\cite{NSR,Melo,Randeria,OhashiGriffin}, we numerically solve the $T_{\rm c}$-equation (\ref{eq.10}), together with the number equation (\ref{eq.9}), to self-consistently determine $T_{\rm c}$ and $\mu(T_{\rm c})$ in the BCS-BEC crossover region. Above $T_{\rm c}$, we only deal with the number equation (\ref{eq.9}) to determine $\mu(T)$, which is used to evaluate the specific heat $C_V$. 
\par
%%%%%%%%%%%%%%%%%%%%%%%%%%%%%%%%%%%%%%%%%%%%%%%%%%%%%%%%%%%%%%%%%%%%%%%%%%%%%
\begin{figure}[t]
\begin{center}
\includegraphics[width=7cm]{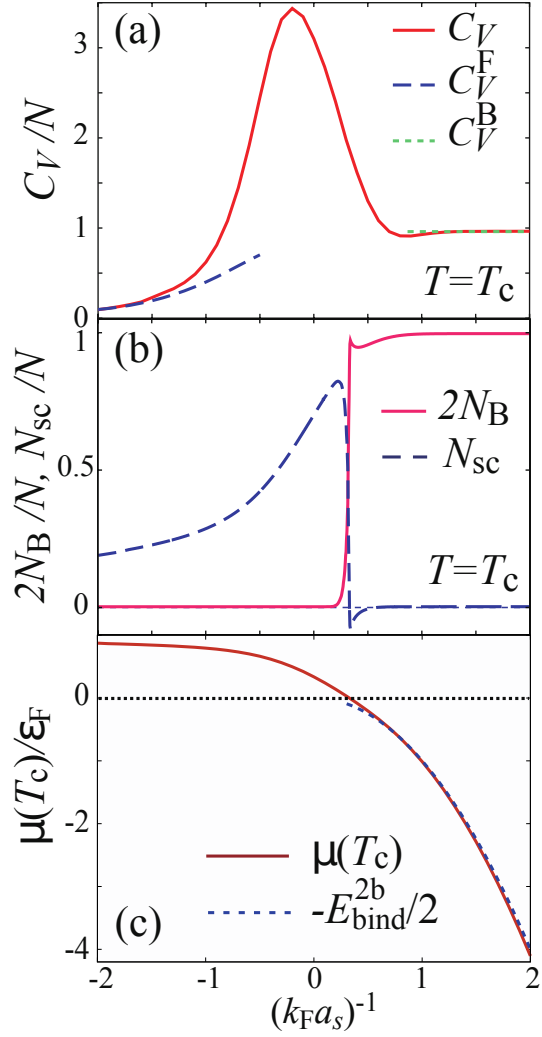}
\end{center}
\caption{(Color online) (a) Calculated specific heat $C_V$ at $T_{\rm c}$, as a function of the interaction strength measured in terms of the inverse scattering length $(k_{\rm F}a_s)^{-1}$, normalized by the Fermi momentum $k_{\rm F}$. We also plot the specific heat $C_V^{\rm F}(T=T_{\rm c})$ in a free Fermi gas, as well as the specific heat $C^{\rm B}_V$ in Eq. (\ref{eq.12}). (b) The number of stable molecules $N_{\rm B}$ at $T_{\rm c}$. $N_{\rm sc}$ is the contribution from scattering states at $T_{\rm c}$. (c) Fermi chemical potential $\mu(T=T_{\rm c})$, normalized by the Fermi energy $\varepsilon_{\rm F}$. The dashed line shows $-E_{\rm bind}^{\rm 2b}/2$, where $E_{\rm bind}=1/(ma_s^2)$ is the binding energy of a two-body bound molecule.}
\label{fig2}
\end{figure}
%%%%%%%%%%%%%%%%%%%%%%%%%%%%%%%%%%%%%%%%%%%%%%%%%%%%%%%%%%%%%%%%%%%%%%%%%%%%%%
\par
\section{Specific heat in the BCS-BEC crossover region above $T_{\rm c}$}
\par
Figure \ref{fig2}(a) shows the specific heat at constant volume $C_V$ in the BCS-BEC crossover regime of an ultracold Fermi gas at $T=T_{\rm c}$. As expected, $C_V$ in the weak-coupling BCS regime ($(k_{\rm F}a_s)^{-1}\lesssim -1$, where $k_{\rm F}$ is the Fermi momentum), as well as that in the strong-coupling BEC regime ($(k_{\rm F}a_s)^{-1}\gesim 1$) are, respectively, well described by the specific heat in a free Fermi gas\cite{Fetter},
\begin{equation}
C_V^{\rm F}(T\ll T_{\rm F})={\pi^2 \over 2}
\left(T \over T_{\rm F}
\right)N
\label{eq.11}
\end{equation}
(where $T_{\rm F}$ is the Fermi temperature), and the specific heat in an ideal Bose gas with $N/2$ molecules at the BEC phase transition temperature $T_{\rm BEC}=0.218T_{\rm F}$\cite{NSR,Randeria,Melo,Phillips}
\begin{equation}
C_V^{\rm B}(T=T_{\rm BEC})=
{15 \over 4}N_{\rm B}
\times
{\zeta(5/2) \over \zeta(3/2)}
=0.963N,
\label{eq.12}
\end{equation}
where $\zeta(3/2)=2.612$ and $\zeta(5/2)=1.341$ are zeta functions. Although $C_V$ continuously changes from $C_V^{\rm F}$ to $C_V^{\rm B}$ in the BCS-BEC crossover, it experiences anomalous enhancement in the unitary regime ($(k_{\rm F}a_s)^{-1}\sim 0$), as seen in Fig. \ref{fig2}(a). 
\par
%%%%%%%%%%%%%%%%%%%%%%%%%%%%%%%%%%%%%%%%%%%%%%%%%%%%%%%%%%%%%%%%%%%%%%%%%%%%%%%
\begin{figure}[t]
\begin{center}
\includegraphics[width=10cm]{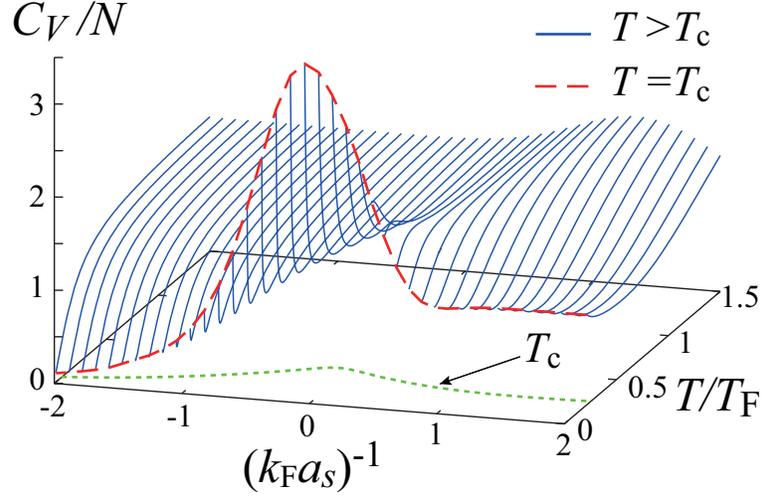}
\end{center}
\caption{(Color online) Calculated specific heat $C_V$, as a function of the temperature in the BCS-BEC crossover regime of an ultracold Fermi gas above $T_{\rm c}$. The dashed line shows the result at $T_{\rm c}$.}
\label{fig3}
\end{figure}
%%%%%%%%%%%%%%%%%%%%%%%%%%%%%%%%%%%%%%%%%%%%%%%%%%%%%%%%%%%%%%%%%%%%%%%%%%%%%%%
\par
This remarkable enhancement of $C_V$ originates from the suppression of the entropy $S=\ln W$ by the appearance of preformed Cooper pairs near $T_{\rm c}$. Since the number of possible micro-states $W$ in a gas of bound molecules with nearly zero center of mass momentum is smaller than $W$ in a simple unbound Fermi gas, the gradual formation of preformed Cooper pairs with decreasing temperature nearing $T_{\rm c}$ suppresses the entropy $S$. When this suppression is more remarkable at lower temperatures, the thermodynamic formula,
\begin{equation}
C_V=T\lef(\frac{\partial S}{\partial T}\rig)_{V,N},
\label{eq.13}
\end{equation}
immediately gives the enhancement of $C_V$. In the unitary regime, such preformed-pair formation occurs near $T_{\rm c}$, so that the amplification of $C_V$ is also restricted to the region near $T_{\rm c}$, as shown Fig. \ref{fig3}.
\par
%%%%%%%%%%%%%%%%%%%%%%%%%%%%%%%%%%%%%%%%%%%%%%%%%%%%%%%%%%%%%%%%%%%%%%%%%%%%%%%
\begin{figure}[t]
\begin{center}
\includegraphics[width=8.0cm]{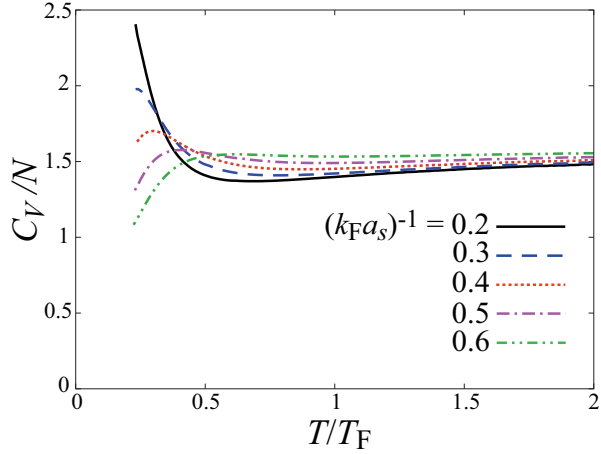}
\end{center}
\caption{(Color online) Specific heat $C_V$ in the BEC side, $0.2\le (k_{\rm F}a_s)^{-1}\le 0.6$.}
\label{fig4}
\end{figure}
%%%%%%%%%%%%%%%%%%%%%%%%%%%%%%%%%%%%%%%%%%%%%%%%%%%%%%%%%%%%%%%%%%%%%%%%%%%%%%%
\par
To see whether these preformed Cooper pairs are stable or fluctuating, it is convenient to divide the NSR correction term $N_{\rm NSR}$ in the number equation (\ref{eq.9}) into the sum of twice the number $N_{\rm B}$ of {\it stable} molecules and the so-called scattering part $N_{\rm sc}$\cite{NSR} involving contribution from fluctuating {\it metastable} preformed pairs\cite{OhashiGriffin,Engelbrecht}. The resulting expression for the total number $N$ of Fermi atoms has the form,
\begin{equation}
N=N_{\rm F}^0+2N_{\rm B}+N_{\rm sc}.
\label{eq.14}
\end{equation}
(For detailed expressions for $N_{\rm B}$ and $N_{\rm sc}$, see the Appendix.) Then, we find in Fig. \ref{fig2}(b) that there are no {\it stable} preformed pairs ($N_{\rm B}=0$) in the unitary regime where $C_V$ is remarkably amplified\cite{note2}, which means that this enhancement is due to the increase of {\it metastable} preformed Cooper pairs or pairing fluctuations. 
\par
Figure \ref{fig2}(b) also indicates that the strong-coupling BEC regime ($(k_{\rm F}a_s)^{-1}\gesim 0.3$) is dominated by long-lived {\it stable} molecules ($N_{\rm B}\simeq N/2$). This naturally explains why $C_V(T=T_{\rm c})$ in this regime is well described by $C_V^{\rm B}$ in Eq. (\ref{eq.12}). As shown in Fig. \ref{fig2}(c), this result is also consistent with the well-known result for the Fermi chemical potential $\mu$ that it becomes negative in the BEC regime, and the magnitude $|\mu|$ approaches half the binding energy $E_{\rm bind}^{\rm 2b}=1/(ma_{\rm s}^2)$ of a two-body bound state in the strong-coupling limit\cite{NSR,Randeria,Melo,OhashiGriffin,Engelbrecht}.
\par
%%%%%%%%%%%%%%%%%%%%%%%%%%%%%%%%%%%%%%%%%%%%%%%%%%%%%%%%%%%%%%%%%%%%%%%%%%%%%%%
\begin{figure}[t]
\begin{center}
\includegraphics[width=15.0cm]{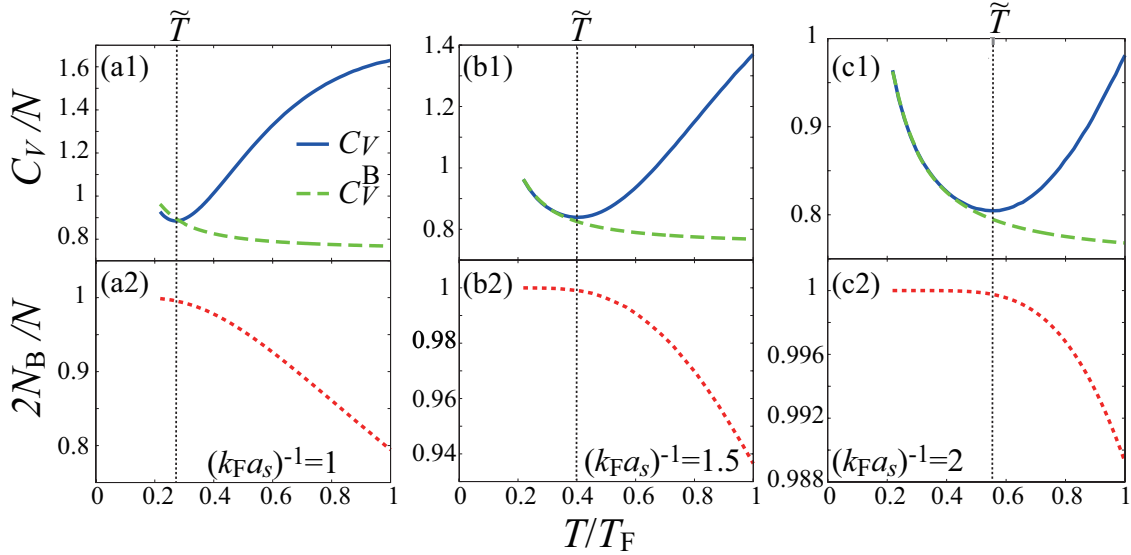}
\end{center}
\caption{(Color online) (a1)-(c1) Specific heat $C_V(T\ge T_{\rm c})$ in the strong-coupling BEC regime ($(k_{\rm F}a_{\rm s})^{-1}\ge 1$). $C_{V}^{\rm B}$ is the specific heat in an ideal gas of $N/2$ bosons with a molecular mass $M=2m$. (a2)-(c2) The number $N_{\rm B}$ of stable pairs. The characteristic temperature ${\tilde T}$ is given as the temperature at which $C_V$ takes a minimal value in the BEC regime.}
\label{fig5}
\end{figure}
%%%%%%%%%%%%%%%%%%%%%%%%%%%%%%%%%%%%%%%%%%%%%%%%%%%%%%%%%%%%%%%%%%%%%%%%%%%%%%%
\par
As seen in Fig. \ref{fig3}, the amplification of $C_V$ in the unitary regime near $T_{\rm c}$ disappears, as one moves to the BEC side ($(k_{\rm F}a_s)^{-1}\gesim 0$). To see this more clearly, we summarize in Fig. \ref{fig4} the temperature dependence of $C_V$ slightly in the BEC side. Noting that long-lived stable molecules appear when $(k_{\rm F}a_s)^{-1}\gesim 0.3$ (see Fig. \ref{fig2}(b)), we expect that the thermal dissociation of these molecules (with a relatively small binding energy $E_{\rm bind}$) is responsible for the temperature dependence of $C_V$ near $T_{\rm c}$ in this regime. Indeed, simply taking into account this effect by dealing with a two-level system with energy $\omega=0$ and $\omega=E_{\rm bind}$, one has
\begin{equation}
C_V=
\left({E_{\rm bind} \over 2T}\right)^2
{\rm sech}^2\left({E_{\rm bind} \over 2T}\right),
\label{eq15}
\end{equation}
which monotonically increases with increasing the temperature when $T\lesssim E_{\rm bind}/2$. This is just the behavior of $C_V(T\gesim T_{\rm c})$ shown in Fig. \ref{fig4} when $(k_{\rm F}a_s)^{-1}\gesim 0.3$, indicating that the increase of $C_{V}$ with increasing the temperature near $T_{\text{c}}$ in this region originates from the thermal dissociation of stable molecules.
\par
We note that the key to understand the reason why the temperature dependence of $C_V$ near $T_{\rm c}$ slightly in the BEC regime is qualitatively different from the that in the unitary regime is the stability of preformed pairs. In the former BEC case where long-lived stable molecules ($N_{\rm B}\simeq N/2$) with a finite binding energy $E_{\rm bind}$ dominate over the system, thermal dissociation of molecules leads to exponential-like temperature dependence of various thermodynamic quantities. Because of this, the entropy $S$ becomes a {\it concave} function of temperature, so that the specific heat $C_V$ given by Eq. (\ref{eq.13}) becomes an increasing function of the temperature. On the other hand, in the case of unitary regime with no stable molecule ($N_{\rm B}=0$), because metastable preformed Cooper pairs are actually pairing fluctuations where formation and dissociation of preformed pairs repeatedly and frequently occur, the binding energy of such a fluctuating quasi-molecule is somehow ambiguous, especially when the molecular lifetime is very short. As a result, fluctuating metastable preformed pairs would not not give an exponential temperature dependence of $S$. However, as mentioned previously, the growth of low-energy pairing fluctuations or metastable preformed pairs with nearly zero center of mass momentum near $T_{\rm c}$ decreases the entropy $S$. In addition, because this growth is more remarkable at lower temperatures near $T_{\rm c}$, the entropy $S$ becomes a {\it convex} function of temperature, so that Eq. (\ref{eq.13}) gives the quite opposite temperature dependence of $C_V$ to the case slightly in the BEC regime ($(k_{\rm F}a_s)^{-1}\gesim 0.3$) near $T_{\rm c}$ .
\par
The above-mentioned difference can be also understood on the viewpoint of the internal energy $E$. In the unitary regime, while the molecular picture is ambiguous, pairing fluctuations are known to induce particle-hole coupling\cite{Tsuchiya2}, leading to a pseudogap structure around the Fermi level. This phenomenon would lower the internal energy $E$ (as in the case of the ordinary BCS state), which would become more remarkable at lower temperature near $T_{\rm c}$, because of the enhancement of pairing fluctuations. As a result, $E$ becomes a convex function of temperature, so that Eq. (\ref{eq.7}) gives the anomalous amplification of $C_V$ in the unitary regime near $T_{\rm c}$. On the other hand, slightly in the BEC regime where low-energy single-particle Fermi excitations near $T_{\rm c}$ are dominated by thermal dissociation of long-lived stable molecules, the internal energy would have an exponential-like temperature dependence, so that Eq. (\ref{eq.7}) gives the increase of $C_V$ with increasing the temperature near $T_{\rm c}$.
\par
Deep inside the BEC regime ($(k_{\rm F}a_s)^{-1}\gesim 0.7$), we see in Figs. \ref{fig5}(a1)-(c1) that the enhancement of $C_V$ is revived near $T_{\rm c}$, although it is not so remarkable as the case of unitary regime. In the temperature region where $C_V$ increases with decreasing the temperature, Figs. \ref{fig5}(a2)-(c2) show that the system is dominated by long-lived {\it stable} pairs ($N_{\rm B}\simeq N/2$). In addition, $C_V$ in this temperature region is well described by the specific heat $C_V^{\rm B}$ of an ideal Bose gas with $N/2$ molecules, as shown in Figs. \ref{fig5}(a1)-(c1). Thus, when one conveniently introduces the characteristic temperature ${\tilde T}$ as the temperature at which $C_V(T)$ takes a minimum value, the region $T_{\rm c}\le T\lesssim {\tilde T}$ may be regarded as an almost ideal Bose gas of long-lived stable molecules. 
\par
We briefly note that such a Bose gas behavior of $C_V$ can be also confirmed analytically. In the strong coupling BEC regime, the particle-particle scattering matrix $\Gamma({\bm q},i\nu_n)$ in Eq.(\ref{eq.5}) is reduced to\cite{Perali}, 
\begin{equation}
\Gamma({\bm q},i\nu_n)=
{8\pi \over m^2a_s}
{1 \over i\nu_n-\frac{q^{2}}{4m}+\mu_{\rm B}},
 \label{eq16}
\end{equation}
where $\mu_{\rm B}=2\mu+E^{\rm 2b}_{\text{bind}}$. In obtaining Eq. (\ref{eq16}), we have used the well-known result in the BEC regime, $\mu\simeq -E^{\rm 2b}_{\rm bind}/2=-1/(2ma_s{\rm s}^2) \ll -\varepsilon_{\rm F}$\cite{NSR,Randeria,Melo}. Substituting Eq. (\ref{eq16}) into the internal energy in Eq.(\ref{eq.8}), one has
\begin{eqnarray}
E=\sum_{\bm q}{q^2 \over 4m}
n_{\rm B}\left({q^2 \over 4m}-\mu_{\rm B}\right)
-E_{\rm bind}{N \over 2}.
\label{eq.17}
\end{eqnarray}
Here, we have assumed that all the Fermi atoms form tightly bound molecules, for simplicity. Since the specific heat $C_V=(\partial E/\partial T)_{V,N}$ is simply obtained from the first term in Eq. (\ref{eq.17}), it is just the same as the specific heat in an ideal gas with $N/2$ two-body bound molecules.
\par
With increasing temperature above ${\tilde T}$, the gradual decrease of the number $N_{\rm B}$ of stable pairs from $N/2$, as shown in Figs. \ref{fig5}(a2)-(c2), indicates the thermal dissociation of molecules. As shown in Eq. (\ref{eq15}), this phenomenon naturally increases $C_V$, giving the deviation from $C_V^{\rm B}$ seen in Figs. \ref{fig5}(a1)-(c1).
\par
%%%%%%%%%%%%%%%%%%%%%%%%%%%%%%%%%%%%%%%%%%%%%%%%%%%%%%%%%%%%%%%%%%%%%%%%%%%%%%%
\begin{figure}[t]
\begin{center}
\includegraphics[width=8cm]{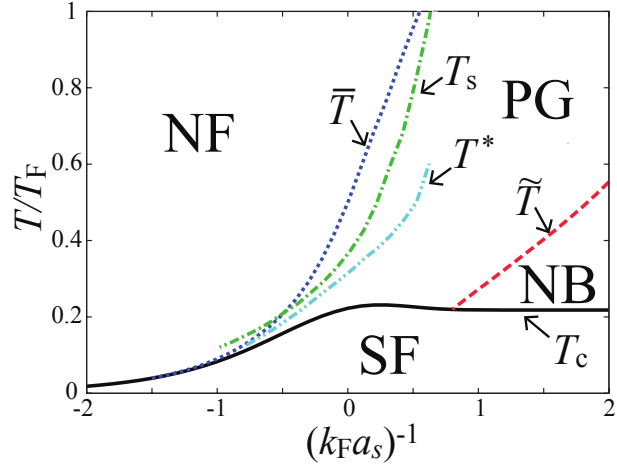}
\end{center}
\caption{(Color online) Phase diagram of an ultracold Fermi gas with respect to the interaction strength $(k_{\rm F}a_s)^{-1}$ and the temperature $T$, scaled by the Fermi temperature $T_{\rm F}$. The characteristic temperature ${\tilde T}$ gives the boundary between the region (NB) of an almost ideal Bose gas with $N/2$ non-condensed long-lived stable pairs and the pseudogap regime (PG), where the system is dominated by metastable preformed Cooper pairs or pairing fluctuations. ${\bar T}$ physically gives the boundary between the normal Fermi gas regime (NF), and PG. The region below $T_{\rm c}$ is in the superfluid state. For comparison, we also plot the previous pseudogap temperature $T^*$\cite{Tsuchiya} obtained from the density of states $\rho(\omega)$, as well as the spin-gap temperature $T_{\rm s}$\cite{Tajima} determined from the spin susceptibility $\chi_{\rm s}$. We note that $T_{\rm c}$ is only the phase transition temperature. ${\tilde T}$, ${\bar T}$, $T^*$, as well as $T_{\rm s}$, are all characteristic temperatures, without being accompanied by any phase transition.}
\label{fig6}
\end{figure}
%%%%%%%%%%%%%%%%%%%%%%%%%%%%%%%%%%%%%%%%%%%%%%%%%%%%%%%%%%%%%%%%%%%%%%%%%%%%%%%
\par
Plotting ${\tilde T}$ in the phase digram of an ultracold Fermi gas in terms of the interaction strength and the temperature, we obtain Fig. \ref{fig6}. As discussed above, this line physically gives the boundary between the region (NB) of an almost ideal Bose gas with $N/2$ non-condensed long-lived stable pairs and the so-called pseudogap regime (PG), where metastable preformed pairs dominate over the system. Particularly at $T_{\rm c}$, this boundary is at $(k_{\rm F}a_s)^{-1}\simeq 0.8$. We birefly point out that this value is consistent with the previous result $(k_{\rm F}a_s)^{-1}\simeq 0.75$\cite{Perali}, which was determined from the analyses of Fermi single-particle excitations.
\par
We note that the boundary between the pseudogap regime (PG) and the normal Bose gas regime (NB) has previously been given by $T'=2|\mu|$ (where $\mu<0$)\cite{Tsuchiya,Tsuchiya2}. The background idea for this characteristic temperature is that $2|\mu|$ eventually coincides with the binding energy $E_{\rm bind}^{2b}=1/(ma_s^2)$ of a two-body bound molecule in the BEC limit, so that stable molecules are expected to appear below $T'\sim E_{\rm bind}^{\rm 2b}$, overwhelming thermal dissociation. However, comparing $T'$ with ${\tilde T}$, one finds that they are actually very different, as $T'\gg {\tilde T}$ (although we do not explicitly show this comparison here). This indicates that, although stable pairs would start to appear around $T'\sim E_{\rm bind}^{2b}$, it does not immediately mean the realization of a molecular Bose gas. To obtain a gas of long-lived stable pairs, we need to further decrease the temperature down to ${\tilde T}$, at least on the viewpoint of the specific heat $C_V$. In this sense, the region between $T'$ and ${\tilde T}$ may be regarded as the crossover region between a gas of metastable quasi-molecules and that of long-lived stable molecules. 
\par
%%%%%%%%%%%%%%%%%%%%%%%%%%%%%%%%%%%%%%%%%%%%%%%%%%%%%%%%%%%%%%%%%%%%%%%%%%%%%%%
\begin{figure}[t]
\begin{center}
\includegraphics[width=7.5cm]{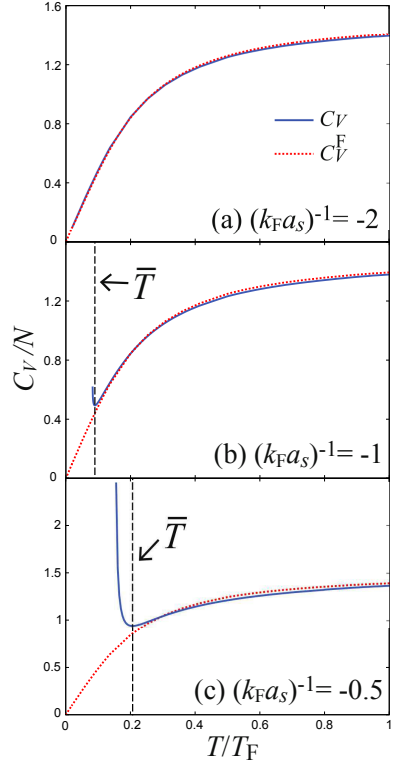}
\end{center}
\caption{(Color online) Calculated specific heat $C_V$ in the BCS side ($(k_{\rm F}a_s)^{-1}\le 0$) as a function of temperature $T$, scaled by the Fermi temperature $T_{\rm F}$. $C_V^{\rm F}$ is the specific heat in a free Fermi gas in Eq. (\ref{eq.14}). The characteristic temperature ${\bar T}$ is determined as the temperature at which $C_V$ becomes minimal.}
\label{fig7}
\end{figure}
%%%%%%%%%%%%%%%%%%%%%%%%%%%%%%%%%%%%%%%%%%%%%%%%%%%%%%%%%%%%%%%%%%%%%%%%%%%%%%%
\par
We note that the physical meaning of ${\tilde T}$ is different from the previous pseudogap temperature $T^*$\cite{Tsuchiya} and the spin-gap temperature $T_{\rm s}$\cite{Kashimura,Tajima}, because the latter two physically give the boundary between the normal Fermi gas regime (NF) and PG. In this regard, we point out that the specific heat $C_V$ can also give the other characteristic temperature, which we denote ${\bar T}$, corresponding to $T^*$ and $T_{\rm s}$. As seen in Fig. \ref{fig7}, when one move to the weak-coupling BCS side ($(k_{\rm F}a_s)^{-1}\lesssim 0$) from the unitary regime, the enhancement of $C_V$ near $T_{\rm c}$ (which is caused by metastable preformed pairs) gradually disappears, and the temperature dependence of $C_V$ is reduces to that in a free Fermi gas, given by
\begin{equation}
C_V^{\rm F}=2\sum_{\bm p}\varepsilon_{\bm p}
{\partial f(\xi_{\bm p}) \over \partial T}.
\label{eq.14}
\end{equation}
Equation (\ref{eq.14}) is proportional to $T$ when $T\ll T_{\rm F}$ (see Eq. (\ref{eq.11})). It approaches the classical Dulong-Petit law, $C_V^{\rm cl}=3N/2$\cite{Fetter} in the high temperature region. Thus, the temperature ($\equiv {\bar T}$) at which $C_V$ takes a minimal value in Fig. \ref{fig7} may be reasonably interpreted as the boundary between the normal Fermi gas regime (NF) and the pseudogap gap regime (PG) dominated by fluctuating metastable preformed Cooper pairs. Indeed, when we plot this characteristic temperature ${\bar T}$ in Fig. \ref{fig6}, it is found to be consistent with $T^*$ and $T_{\rm s}$, as expected.
\par
%%%%%%%%%%%%%%%%%%%%%%%%%%%%%%%%%%%%%%%%%%%%%%%%%%%%%%%%%%%%%%%%%%%%%%%%%%%%%%%
\begin{figure}[t]
 \begin{center}
  \includegraphics[width=8cm]{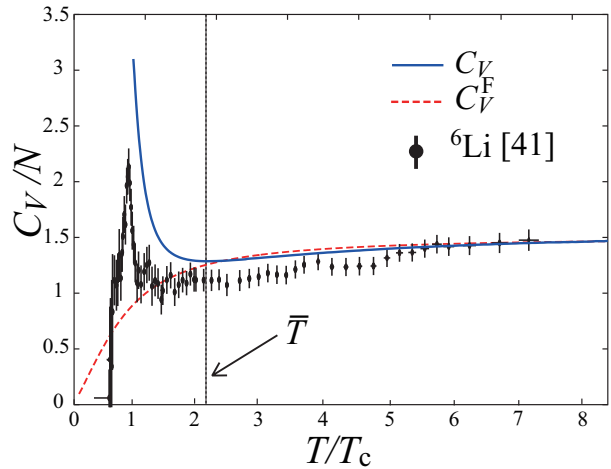}
 \end{center}
 \caption{(Color online) Comparison of our result with the resent experiment on a $^6$Li unitary Fermi gas\cite{Ku}. In this figure, the temperature is normalized by $T_{\rm c}$. $C_V^F$ is the specific heat in a free Fermi gas.}
 \label{fig8}
\end{figure}
%%%%%%%%%%%%%%%%%%%%%%%%%%%%%%%%%%%%%%%%%%%%%%%%%%%%%%%%%%%%%%%%%%%%%%%%%%%%%%%
\par
Finally, we compare our result with the recent experiment on a $^6$Li unitary Fermi gas\cite{Ku}. Figure \ref{fig8} shows that our result well explains the observed amplification of $C_V$ near $T_{\rm c}$, indicating that the observed anomaly is due to {\it metastable} preformed Cooper pairs. However, Fig. \ref{fig8} also shows that our result overestimates this enhancement near $T_{\text{c}}$. In this regard, we recall that a finite spacial resolution inherent in this experiment in a trapped geometry could lead to a possible suppression of the specific heat near $T_{\rm c}$\cite{Ku}. We also point out that, since we deal with pairing fluctuations within the simplest NSR level, inclusion of higher-order strong-coupling corrections beyond this approximation may also be important to correctly describe the behavior of the specific heat $C_V$, especially near $T_{\rm c}$. Thus, we need further analyses to quantitatively explain this experimental result. 
\par
%%%%%%%%%%%%%%%%%%%%%%%%%%%%%%%%%%%%%%%%%%%%%%%%%%%%%%%%%%%%%%%%%%%%%%%%%%%%%
\section{Summary}
\par
To summarise, we have discussed the specific heat at constant volume $C_V$ in the BCS-BEC crossover regime of an ultracold cold Fermi gas. Including pairing fluctuations within the framework of the strong-coupling theory developed by Nozi\`eres and Schmitt-Rink, we clarified the temperature dependence of this thermodynamic quantity over the entire BCS-BEC crossover region above $T_{\rm c}$. In the unitary regime, we found that the specific heat is anomalously amplified near $T_{\rm c}$, which is due to the appearance of fluctuating {\it metastable} preformed Cooper pairs. Although this anomaly once disappears as one goes to the BEC side, $C_V$ was found to be again enhanced near $T_{\rm c}$ with further increasing the interaction strength. We showed that this regime is dominated by long-lived {\it stable} pairs, and the enhancement of $C_V$ in this region agrees well with the case of an ideal molecular Bose gas. 
\par
Using these results, we determined the characteristic temperature ${\tilde T}$ which physically distinguish between the pseudogap regime where the system is dominated by metastable preformed Cooper pairs (or pairing fluctuations) and the region of an almost ideal Bose gas with $N/2$ non-condensed stable pairs. 
\par
From the temperature dependence of the specific heat in the BCS side, we also determined the other characteristic temperature ${\bar T}$, which physically distinguish between the normal Fermi gas regime and the pseudogap regime. Using ${\tilde T}$ and ${\bar T}$, as well as the superfluid phase transition temperature $T_{\rm c}$, we obtained the phase diagram of an ultracold Fermi gas in terms of the interaction strength and the temperature, consisting of (1) the normal Fermi gas regime, (2) the pseudogap regime dominated by metastable preformed Cooper pairs or pairing fluctuations, (3) the region of an almost ideal Bose gas with $N/2$ non-condensed long-lived stable pairs, and (4) the superfluid phase below $T_{\rm c}$. Although ${\tilde T}$ and ${\bar T}$ are not accompanied by any phase transition, they are still useful in considering the strong-coupling properties of an ultracold Fermi gas in the BCS-BEC crossover region.
\par
We note that, although the background physics of ${\bar T}$ is similar to that of the previous pseudogap temperature $T^*$, which is determined from the density of states $\rho(\omega)$, as well as that of the spin-gap temperature $T_{\rm s}$, which is determined from the spin susceptibility $\chi_{\rm s}$, it is difficult to obtain the characteristic temperature corresponding to ${\tilde T}$ from $\rho(\omega)$ and $\chi_{\rm s}$. This is because they almost vanish in the strong-coupling BEC regime, due to the formation of tightly bound spin-singlet pairs with a large binding energy. In contrast, the specific heat is not suppressed in the BEC regime, so that we can safely determine ${\tilde T}$ to identify the region consisting of stable pairs below ${\tilde T}$. In addition, the specific heat is known to exhibit singularity at $T_{\rm c}$. These advantages indicate that the specific heat is a useful quantity in constructing the phase diagram of an ultracold Fermi gas in the BCS-BEC crossover region.
\par
We note that we have included strong-coupling corrections within the simplest NSR theory in this paper. In this regard, while the NSR theory can describe the BCS-BEC crossover behavior of $T_{\rm c}$, this strong-coupling theory is known to overestimate the pseudogap phenomenon associated with pairing fluctuations\cite{Tsuchiya,Kashimura}. Since the NSR specific heat at the unitarity overestimates the observed enhancement of $C_V$ near $T_{\rm c}$ in a $^6$Li Fermi gas (see Fig. \ref{fig8}), a more sophisticated treatment of pairing fluctuations beyond the NSR theory would be necessary, in order to quantitatively explain this experiment. In addition, since the NSR theory completely ignores an effecvtive interaction between molecular bosons\cite{StrinatiB,Petrov}, it is also a crucial issue to clarify to what extent this molecular interaction affects the characteristic temperature ${\tilde T}$ (which physically gives the boundary between the region of (long-lived) stabe molecules and the region of metastable preformed pairs). For this problem, the so-called self-consistent $T$-matrix approximation\cite{Haussmann} would be useful. 
\par
We also note that we have only dealt with the normal state above $T_{\rm c}$ in this paper. Thus, extension of the present theory to the superfluid phase below $T_{\rm c}$ is also an interesting challenge. In addition, we have ignored effects of a harmonic trap in this paper. Although these effects should in principle be unimportant regarding the fact that the recent experimental result shown in Fig.\ref{fig8} represents that of a uniform Fermi gas, it has been pointed by the authors of this experiment\cite{Ku} that the trap geometry may induce an error in the temperature measurement. To quantitatively compare our result with the experiment data, we need to theoretically include this point. Although the pseudogap phenomenon associated with pairing fluctuations has recently attracted much attention in cold Fermi gas physics, the pseudogap temperature between the normal Fermi gas regime and the pseudogap regime has so far been mainly discussed. Thus, our results would contribute to the further understanding of BCS-BEC crossover physics in an ultracold Fermi gas on the viewpoint of the preformed pairs.
\par
%%%%%%%%%%%%%%%%%%%%%%%%%%%%%%%%%%%%%%%%%%%%%%%%%%%%%%%%%%%%%%%%%%%%%%%%%%%%%%%
\par
\acknowledgements
We would like to thank D. Inotani and M. Matsumoto for useful discussions. This work was supported by KiPAS project in Keio University. R.H. and H.T. was supported be a Grant-in-Aid for JSPS fellows. YO was also supported by Grant-in-Aid for Scientific research from MEXT and JSPS in Japan (No.25400418, No.15H00840).
\par
%%%%%%%%%%%%%%%%%%%%%%%%%%%%%%%%%%%%%%%%%%%%%%%%%%%%%%%%%%%%%%%%%%%%%%%%%%%%%%%
%  Appendix
%%%%%%%%%%%%%%%%%%%%%%%%%%%%%%%%%%%%%%%%%%%%%%%%%%%%%%%%%%%%%%%%%%%%%%%%%%%%%%%
\appendix
\section{Expression for the number $N_{\rm B}$ of stable molecules}
\par
To extract the contribution of stable molecules from the NSR term $N_{\rm NSR}$ in Eq. (\ref{eq.9}), it is convenient to write it in the spectral representation, as
\begin{equation}
N_{\rm NSR}=2\int_{-\infty}^\infty d\omega
n_{\rm B}(\omega)\rho_{\rm B}(\omega).
\label{eq.a1}
\end{equation}
Here, $\rho_{\rm B}(\omega)=\sum_{\bm q}A_{\rm B}({\bm q},\omega)$ may be viewed as the molecular density of states, and the factor two means that each molecule consists of two Fermi atoms. The molecular spectral weight $A_{\rm B}(\omega)$ in $\rho_{\rm B}(\omega)$ has the form,
\begin{equation}
A_{\rm B}({\bm q},\omega)=
-{1 \over \pi}
{\rm Im}
\left[
\Gamma({\bm q},\omega_+)
{\partial \over \partial (2\mu)}\Pi({\bm q},\omega_+)
\right].
\label{eq.a2}
\end{equation}
In Eq. (\ref{eq.a2}), we have used the simplified notations, $\Gamma({\bm q},\omega_+)=\Gamma({\bm q},i\nu_n\to\omega+i\delta)$, and $\Pi({\bm q},\omega_+)=\Pi({\bm q},i\nu_n\to\omega+i\delta)$, where $\delta$ is an infinitesimally small positive number. When the analytic continued particle-particle scattering matrix $\Gamma({\bm q},\omega_+)$ has a real pole at $\omega=\omega_{\bm q}$, it can be approximated to
\begin{eqnarray}
\Gamma({\bm q},\omega_+)=
&=&{1 \over \displaystyle
{m \over 4\pi a_s}+\Pi({\bm q},\omega_+)-\sum_{\bm p}{1 \over 2\vep_{\bm p}}
}
\nonumber
\\
&\simeq&
{1 \over [\omega_+-\omega_{\bm q}]
{\partial \over \partial\omega_{\bm q}}
\Pi({\bm q},\omega_{\bm q})
}.
\label{eq.a3} 
\end{eqnarray}
The contribution of the pole at $\omega=\omega_{\bm q}$ to the number equation is evaluated by substituting Eq. (\ref{eq.a3}) into Eq. (\ref{eq.a2}). Since the real pole $\omega_{\bm q}$ physically describes the dispersion of a stable molecule,
\begin{equation}
N_{\rm B}
=\sum_{{\bm q}:{\rm pole}}
n_{\rm B}(\omega_{\bm q})
{
{\partial \over \partial(2\mu)}\Pi({\bm q},\omega_{\bm q})
\over
{\partial \over \partial\omega_{\bm q}}\Pi({\bm q},\omega_{\bm q})
}
\label{eq.a4}
\end{equation}
has the meaning of the number of stable pairs, where the summation is taken over real poles of $\Gamma({\bm q},\omega_+)$. 
\par
The contribution $N_{\rm sc}$ of scattering states to the number $N$ of Fermi atoms is then given by
\begin{equation}
N_{\rm sc}=N_{\rm NSR}-2N_{\rm B}.
\label{eq.a5} 
\end{equation}
\par

%%%%%%%%%%%%%%%%%%%%%%%%%%%%%%%%%%%%%%%%%%%%%%%%%%%%%%%%%%%%%%%%%%%%%%%%%%%%%%%
\par


\begin{thebibliography}{100}
\bibitem{note2} We see in Fig. \ref{fig2}(b) that $N_{\rm sc}$ is slightly negative when $(k_{\rm F}a_{\rm s})^{-1}\gesim 0.3$. This is because all the scattering states cannot actually be simply described by quasi-molecules with a finite lifetime\cite{OhashiGriffin,Engelbrecht}.
\end{thebibliography}
\end{document}